# YOCO: A Hybrid In-Memory Computing Architecture with 8-bit Sub-PetaOps/W In-Situ Multiply Arithmetic for Large-Scale AI

Zihao Xuan[1], Yuxuan Yang[2], Wei Xuan[1], Zijia Su[2], Song Chen[2], Yi Kang[2*]
[1]The Hong Kong University of Science and Technology, Hong Kong SAR, China
[2]School of Microelectronics, University of Science and Technology of China, Hefei, China
*Corresponding author: ykang@ustc.edu.cn

***Abstract*—In this paper, we further explore the potential of analog in-memory computing (AiMC) and introduce an innovative artificial intelligence (AI) accelerator architecture named YOCO, featuring three key proposals: (1) YOCO proposes a novel 8-bit in-situ multiply arithmetic (IMA) achieving 123.8 TOPS/W energy-efficiency and 34.9 TOPS throughput through efficient charge-domain computation and time-domain accumulation mechanism. (2) YOCO employs a hybrid ReRAM-SRAM memory structure to balance computational efficiency and storage density. (3) YOCO tailors an IMC-friendly attention computing flow with an efficient pipeline to accelerate the inference of transformer-based AI models. Compared to three SOTA baselines, YOCO on average improves energy efficiency by up to 3.9×~19.9× and throughput by up to 6.8×~33.6× across 10 CNN/transformer models.**

## I. Introduction

In recent years, artificial intelligence (AI) algorithms, exemplified by convolutional neural networks (CNNs) and large language models (LLMs), have exhibited tremendous potential to revolutionize natural language processing (NLP) and computer vision tasks [1]. Driven by the scaling laws of AI, the demand for powerful computing hardware has increased dramatically to ensure the efficient execution of these algorithms [2]. For now, various high-performance AI accelerators, such as GPUs and TPUs, have been developed to meet this demand. However, these devices, based on von Neumann architecture, face significant energy efficiency bottlenecks due to "transistor scaling" limitation and the "memory wall" problem, particularly in the context of large-scale AI.

In-memory computing (IMC) technology has emerged as a promising solution by merging memory and computation units, thereby effectively reducing the cost of data movement and maximizing both bandwidth and energy-efficiency of computing [3]. This approach shows great potential for addressing the inefficiencies of traditional accelerators, especially for memory-bound tasks. Nevertheless, applying IMC to accelerate the inference of large-scale AI models presents several specific challenges, particularly given the diverse requirements of different models.

One primary challenge arises from costly analog-to-digital converters and digital-to-analog converters (ADCs/DACs). IMC architectures typically adopt in-situ arithmetic, where multiply-accumulate (MAC) operations are performed in analog form. This process requires DACs to generate analog inputs and ADCs to produce digital outputs. However, these ADCs and DACs are energy-intensive, consuming up to 85% of power in architectures like ISAAC [4], limiting system efficiency.

Additionally, most IMCs rely on a single type of memory, which restricts their flexibility to handle both static and dynamic matrix computations commonly found in transformer models [5]. For example, SRAM-based IMCs provide high computational speed but are constrained by limited memory size, making them inadequate for storing large-scale AI models. In contrast, ReRAM-based IMCs offer high-density storage but suffer from low-endurance and energy-intensive write operations, which hampers their ability to efficiently execute dynamic matrix computations.

Moreover, neural network algorithms often include complex computational patterns, such as *softmax* and *normalization*, which further constrain IMC performance. It is essential to account for these patterns when designing efficient computing dataflows to fully unlock the potential of IMC.

To address these challenges, this paper highlights the opportunities in analog computing, hybrid memory design, and pipelined computing to achieve energy-efficient acceleration of large-scale AI inference. The main contributions of this paper are as follows:

- We propose a novel in-charge computing mechanism to push *input data conversion, bit-wise multiplication, parallel accumulation, and weighted summation* all-in analog domain for energy-saving multi-bit MAC with low precision loss.
- We introduce an in time-domain accumulating approach to increase the size of the analog computing array with *high signal margin*, further reducing the energy consumption of both ADCs and partial sum (*Psum*) movements.
- We employ a hybrid ReRAM-SRAM memory design to trade-off computational flexibility and storage density. This design supports both *dynamic* and *static* matrix multiplication. Additionally, we design an IMC-friendly attention computing pipeline to accelerate transformer models.
- We develop a new AiMC architecture, named YOCO, with a hierarchical structure spanning *Array, IMA, Tile, and Chip* levels. YOCO integrates the aforementioned key features to enhance computing flexibility, efficiency, and performance.

Finally, we conducted a comprehensive evaluation of the YOCO architecture. At the circuit level, YOCO's in-situ multiply arithmetic (IMA) achieves an average energy efficiency of 123.8 TOPS/W, a throughput of 34.9 TOPS, and a computation error of less than 0.98%, for 8-bit 1024×256 vector-matrix multiplications (VMM). At the architecture level, YOCO outperforms three baseline models across 10 benchmarks, improving energy efficiency by 3.9× to 19.9× and throughput by 6.8× to 33.6×.

## II. Background And Motivation

### A. DNNs and Transformer Models

Deep Neural Networks (DNNs) are a class of advanced AI algorithms characterized by their deep layered architectures, which consist of multiple artificial neurons that enable the capture of intricate patterns in data. As DNNs have evolved, various types of network layers have emerged, including fully connected layers, convolutional layers, attention layers, and more. These layers share a fundamental requirement: matrix computation is the core operation. A prominent example of this is the widely used attention computing in transformer (see Fig. 1(a)), which can be represented by the following equation [5]:

$$\text{Attention}(Q,K,V)=\text{Softmax}\left(\frac{QK^T}{\sqrt{d_k}}\right)V \tag{1}$$

Certainly, attention computation is resource-intensive, requiring high memory bandwidth and numerous matrix multiplications (over 90%), especially in large-language models (LLMs) like GPT and BERT, which may contain billions of parameters.

Data reuse is crucial for optimizing attention computation. Effective data reuse minimizes memory access and computation redundancy, thereby improving both speed and energy efficiency. Techniques such as caching frequently accessed matrices, reusing KV matrices across multiple attention heads, and optimizing matrix multiplication dataflow are commonly employed. In IMC, data reuse is important to overcome limited memory bandwidth, particularly in handling the high volume of intermediate data generated during attention calculations.

### B. Analog Processing-in-Memory Basic

The scaling law of DNN necessitates the urgent development of hardware acceleration systems with high computational efficiency to address the constraints of limited energy resources. Leveraging the inherent tolerance of DNNs to computational noise, there is increasing emphasis on utilizing energy-efficient AiMC to enhance matrix multiplication performance, particularly for DNN inference. Fig. 1(b) demonstrates the basic principle of AiMC for $QK^T$ computing. The transpose of $K$ matrix is pre-stored in the memory array, the $Q$ vector ($q_0$, $q_1$, …, $q_n$) is encoded into analog signals by DACs and propagated horizontally along the word lines of array, parallel analog MACs are performed within memory and compute cells (MCCs), and the vertically accumulated results are summed and converted into digital output by ADCs. The dual-function MCC, combining both memory and compute, eliminates weight movement during MAC operations and facilitates AiMC achieve maximum spatial parallelism and minimal time waste.

### C. ADCs/DACs Cost Analysis

In a classic AiMC architecture, ADCs and DACs dominate energy consumption, limiting the advantages of AiMC (see Fig. 1(c)). According to REALLA's analysis [6], the high overhead of ADCs/DACs is primarily due to computation performed in bit-slice and block-wise modes. In bit slicing, inputs are processed over multiple cycles, weights across multiple parallel columns of array, and fully multi-bit MACs are realized digitally. The number of ADC conversions per MAC (converts/MAC) is proportional to the product of the input and weight slices. Some works suggest that increased bit-slice enables the use of lower-resolution data converters. However, more input slices reduce DAC overhead but significantly increase ADC costs, and additional weight slices do not guarantee lower-resolution ADC. Shifting and adding in the digital domain amplify analog computation and quantization errors. In block-wise computing, weight matrix is divided into smaller blocks to match the size of analog computing array. The converts/MAC are also related to the number of blocks. Clearly, more bit slices and computing blocks mean more converts/MAC and results in more energy consumption.

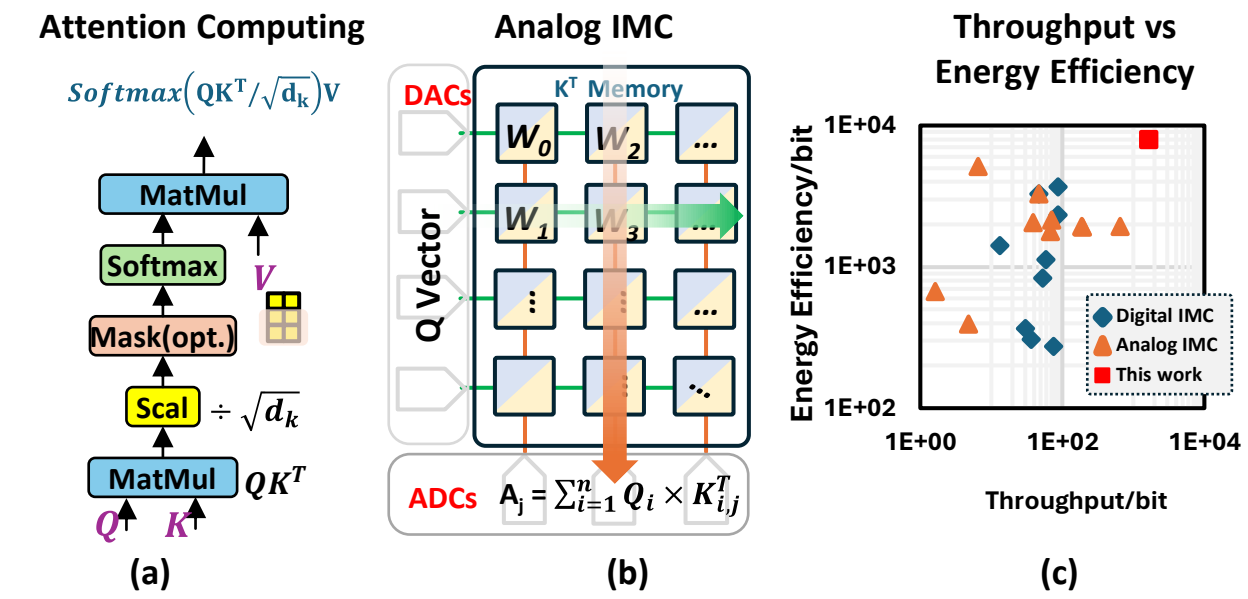

Fig. 1 (a) Scaled Dot-Product Attention. (b) AiMC topology for $QK^T$. (c) Normalized throughput versus energy efficiency of recent IMCs.

Table I. ADCs/DACs Cost Comparison

| Architecture | Slice Weight | Slice Input | Block Size | ADC Cost | DAC Cost | Memory Type | Accuracy Loss |
|---|---|---|---|---|---|---|---|
| ISAAC [4] | Yes | Yes | Small | High | Low | ReRAM | High |
| RAELLA [6] | Yes | Yes | Mid | High | Low | ReRAM | Low |
| TIMELY [7] | Yes | No | Large | Low | Low | ReRAM | High |
| C-Ladder [8] | Yes | No | Small | High | High | DRAM | Low |
| C-2C [9] | No | No | Small | Low | High | SRAM | Low |
| Our (YOCO) | No | No | Large | Low | Low | Hybrid | Low |

### D. Motivation

Some prior works have explored various bit-slice and computing block size to alleviate the cost of ADCs/DACs. Table I highlights the performance gap in recent designs [4, 6-9]. For now, modern AiMC has not yet reached its full potential. As DNNs evolve into the era of LLM, the dual demands for storage density and computational efficiency have intensified, exploring heterogeneous storage and pipeline computing in IMC is beneficial for the computation of large-scale models.

## III. YOCO: You Only Charge Once

We first discuss the principle of the proposed in-charge computing (see sections III-A), followed by design of the time-domain accumulator (see section III-B). Next, we present the overall YOCO architecture in sections III-C, and finally, we introduce our pipeline design in Section III-D.

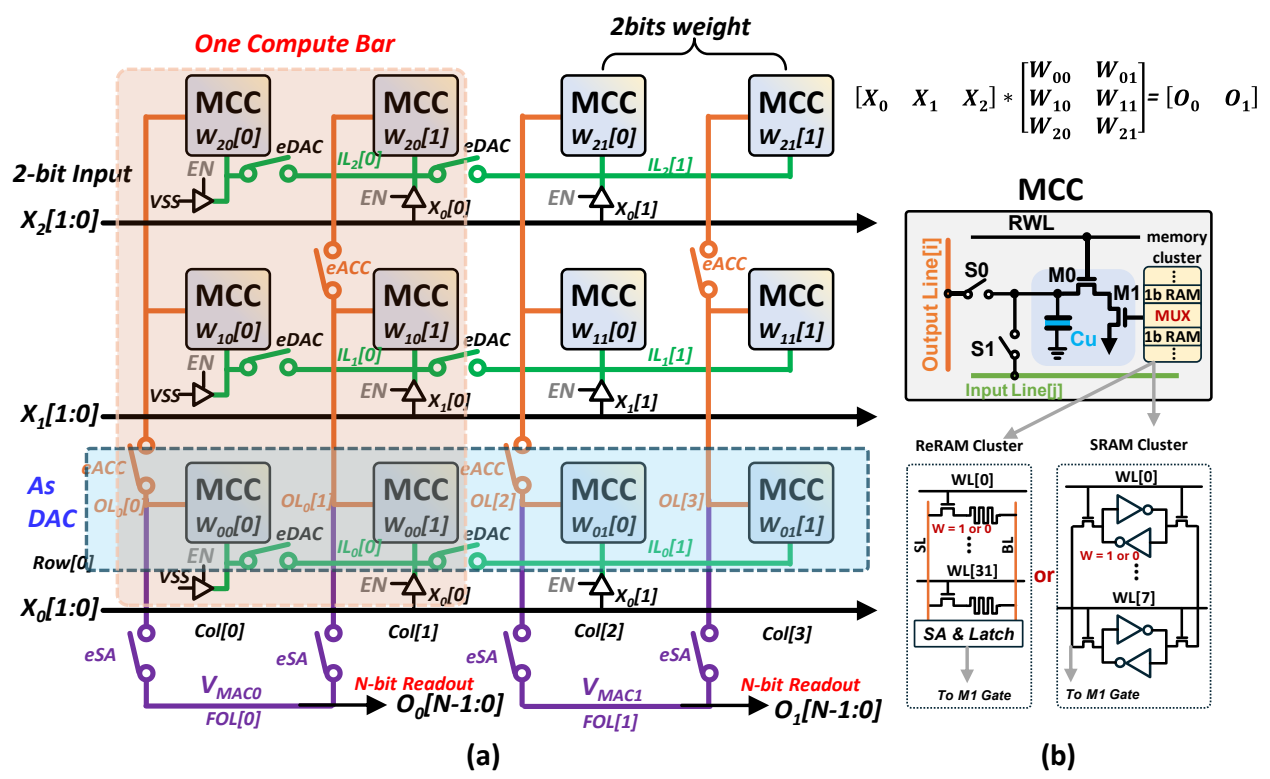

Fig. 2. (a) An example of AiMC array design for 3×2 VMM with 2bits input and weights. (b) MCC with unit capacitor and memory cluster.

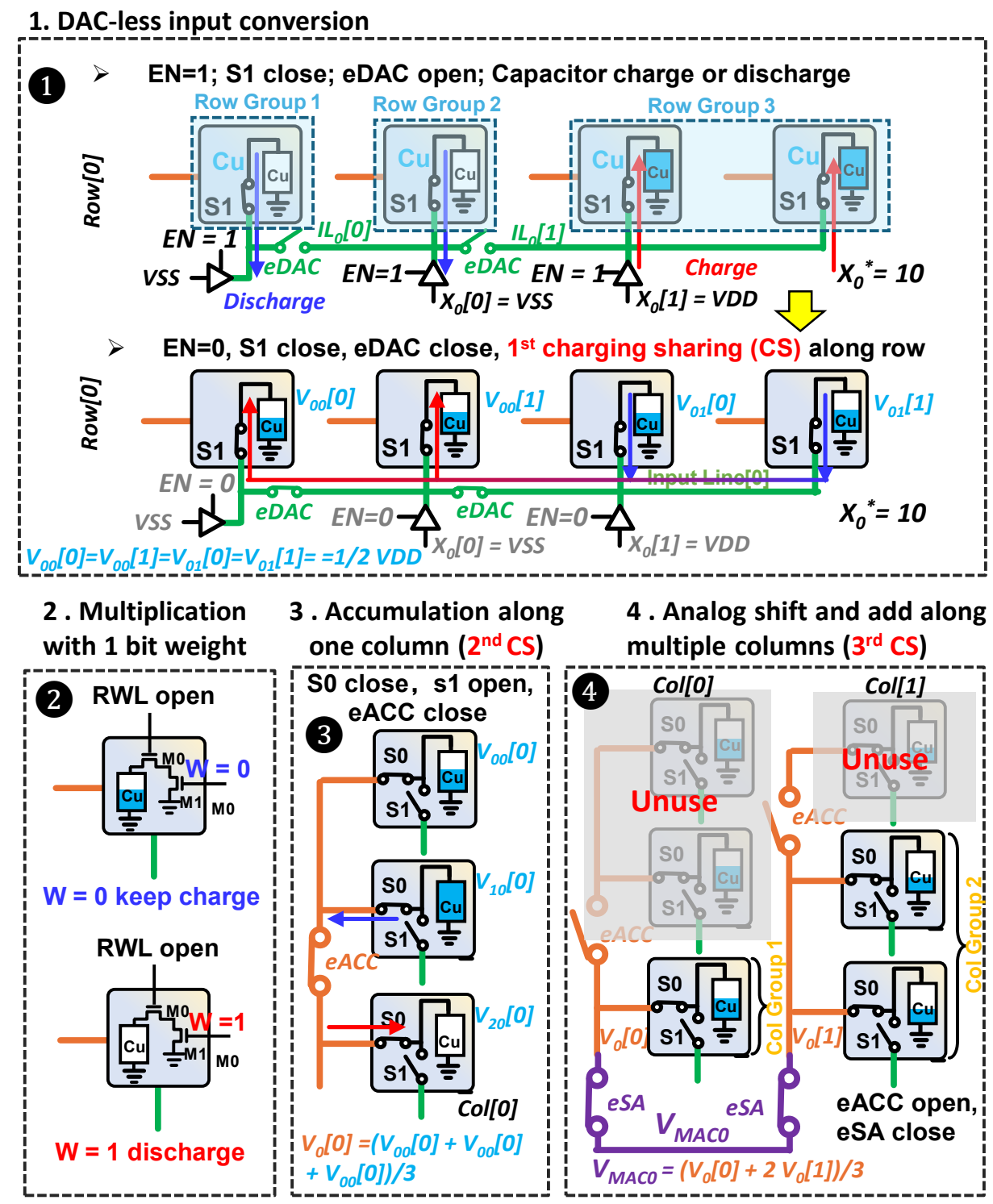

Fig. 3. Four times charge-sharing for in-situ multi-bit computing.

## A. Principle of Proposed in-Charge Computing

In a traditional charge-domain IMC circuit, each memory and computing cell (MCC) is equipped with a unit capacitor (see in Fig. 2(b)) solely for parallel accumulation. Therefore, we repurpose these unit capacitors to perform various functions through multiple charge sharing (MCS), including input data conversion, bitwise multiplication, parallel accumulation, and weighted summation. This enables fully analog multi-bit computation without input and weight slices. Fig. 2(a) shows a simple schematic of the proposed in-charge computing array, featuring 3×4 MCCs to support full parallel 3×2 VMM operations with 2-bit inputs and weights. In our simplified array, low-cost switches such as *eDAC*, *eACC*, and *eSA* are employed to reorganize the MCCs. In each row, MCCs are grouped in a 1:1:2 ratio through two eDAC switches, allowing the unit capacitors in the same row to form a 2-bit DAC for parallel 2-bit input data conversion. In the column direction, each MCC can perform only 1-bit weighted multiplication at a time. To facilitate 2-bit weight computations, two columns of MCCs are organized as a "compute bar" (CB). In each CB, *eACC* switches group the MCCs in each column, according to a unique configuration: the MCCs in the first column are grouped in a 1:2 ratio, while those in the second column are grouped in a 2:1 ratio. As a result, in a CB, only one MCC from the first column is connected to the final output line (FOL) via *eSA*, while two MCCs from the second column are connected. This establishes a 1:2 ratio in the number of unit capacitors between two columns involved in the final charge sharing, thereby inherently achieving in-situ shift and add for weighted summation.

Through additional switches, unit capacitors within MCCs can be dynamically reconfigured and grouped to perform different tasks by directing charge-sharing along specified paths, color-coded in the diagram to indicate stages of operation—green for DAC, orange for analog accumulation, and purple for weighted summation. Fig. 3 provides examples of specific switch configurations for each function, highlighting the array's flexibility for various analog computation needs. It includes four phases: **(1) DAC-Less Input Conversion by Row Charge Sharing:** When the *eDAC* switches are open and the tri-state gate is enabled (*EN=1*), only the first group's input is fixed at $V_{SS}$, while the capacitors inside the other groups charge or discharge according to the input binary code $X_i[n]$, here $i$ is row index and $n$ is bit index. After this step, the tri-state gate is disabled (*EN=0*), and *eDAC* switches are closed, all capacitors in each row to charge sharing until reach equal voltage. If we set $X_0$ [1:0] is "10", the input conversion voltage is 1/2 $V_{DD}$ (see ❶ in Fig. 3). The row capacitors are naturally organized as DAC via *eDAC* switches, and each row of the array requires only additional N tri-state gates and switches, resulting in negligible area and power consumption. **(2) Multiplication with 1-bit Weight.** In an MCC, analog multiplication is performed by two transistors *M0* and *M1*, whose gates are controlled by read word line (RWL) and memory cell, respectively. When RWL is activated, each MCC individual determines whether to release the charge of the capacitor based on 1-bit weight ("1" means keep and "0" means released). The final voltage within capacitor represents the result of multiplying between N-bit input and 1-bit weight (see ❷ in Fig. 3). **(3) Parallel Accumulation by Column Charge Sharing.** After the multiplication is completed, the S0 and *eACC* switches are closed (see ❸ in Fig. 3), allowing the capacitors of the MCCs in the same column to share charges, enabling full parallel accumulation. **(4) Weighted Summation by Multi-Column Charge Sharing.** After the completion of the third phase, each column's output represents an analog MAC result based on 1-bit weight. To achieve multi-bit weight MAC, as shown in ❹ of Fig. 3, we control the *eACC* switches to open and the *eSA* to close, so that only a unit capacitor in the first column participates in charge sharing, while two participate in the second column do so. The final shared voltage represents the bit-shifted sum of the two-column voltages, averaged accordingly.

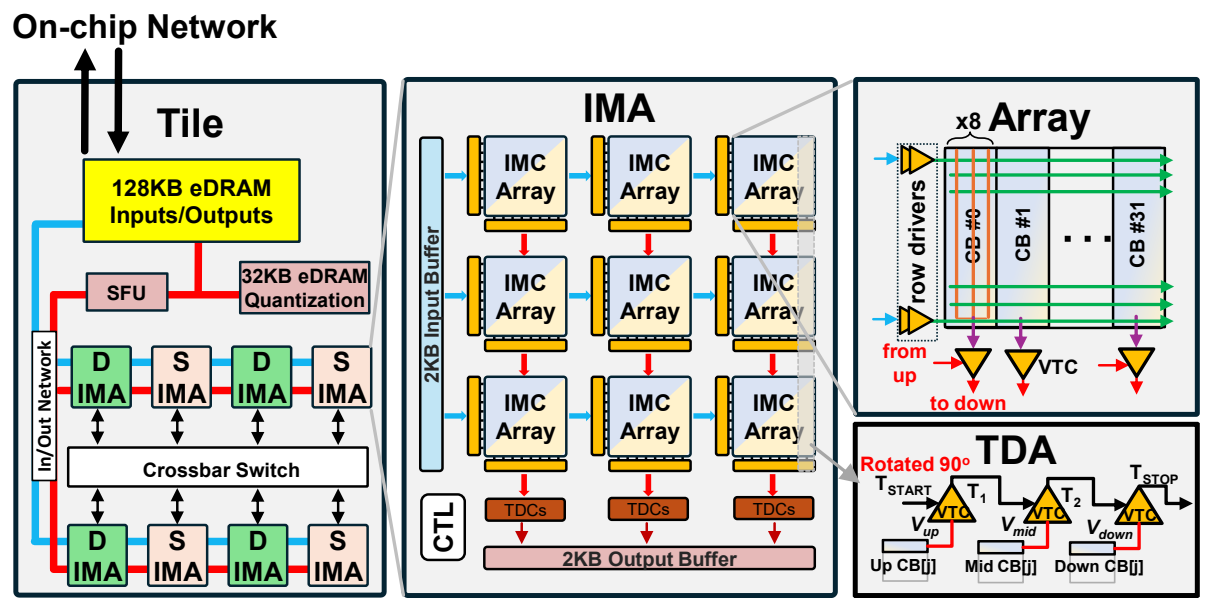

Fig. 4. Hierarchical YOCO architecture with in-charge computing array, time-domain IMA, and hybrid-memory based tile.

Overall, our design showcases a highly flexible and power-efficient approach for analog computation, enabling multiple functions to be performed sequentially by configuring the charge transfer pathways within each MCC, and the fully multi-bit computing process only requires charging once.

### *B. Time-Domain Accumulation*

In our array, each CB generates an MAC voltage, which represents an analog partial sum. We can utilize time-domain accumulators (TDAs) to vertically accumulate these partial sums. A TDA (see the right bottom in Fig. 4) consists of serial head-to-tail voltage-to-time converters (VTCs), where the voltages from multiple CBs are converted into time signal and accumulated in the time-domain with high signal margin. When the previous VTC completes a signal transition, it releases a pulse signal to trigger the next VTC. The start signal is input from the top VTC, and the stop signal is collected from the bottom VTC. Ultimately, the time-to-digital converter (TDC) converts the time difference between the start and stop signal into digital output. To account for the inherent delay in the circuit, we use a redundant column of CBs that can be shared across the entire macro as a reference. The output of this reference is then fed into the TDC as the actual start signal.

### *C. Hierarchical YOCO Architecture*

**Memory and Compute Cell and Array:** As shown in Fig. 3(b), each MCC includes a unit capacitor *Cu*, two switches *S1* and *S0*, two transistors *M0* and *M1* (as an analog 1-bit multiplier), and a memory cluster containing multiple 1-bit RAMs and a multiplexer (MUX). The 1-bit RAM can be implemented with either SRAM for performance-prioritized computation or 1b ReRAM for storage-density-prioritized computation. For 8-bit input resolution, each row of array contains at least 256 MCCs, divided into 9 groups by an *eDAC* with ratios of 1: 1: 2: …: 128. For 8-bit weight, each column includes at least 128 MCCs, with every 8 columns forming a CB. Within each CB, the output lines are split via *eACC* and *eSA* with ratios of 1: 2: …: 128. The array performs fully parallel VMM and each CB's output voltage representing the dot product of the weight and input vectors.

**In-Situ Multiply Accumulate:** To reduce data movement and minimize ADC usage, we integrate AiMC arrays in local domain to form an in-situ multiply accumulate (IMA) (Fig. 4). Arrays are interconnected directly through row drivers and column TDAs. Inputs are multicast across arrays horizontally via row drivers (indicated by blue arrows), while partial sum results from multiple arrays are aggregated vertically through TDAs (red arrows) and read out by an 8-bit TDC. Each array is controlled by power gating, allowing the computational scale of IMA to be reconfigurable and energy-saving. Both the input and output of IMA are equipped with 2KB buffers for temporary data storage and reuse.

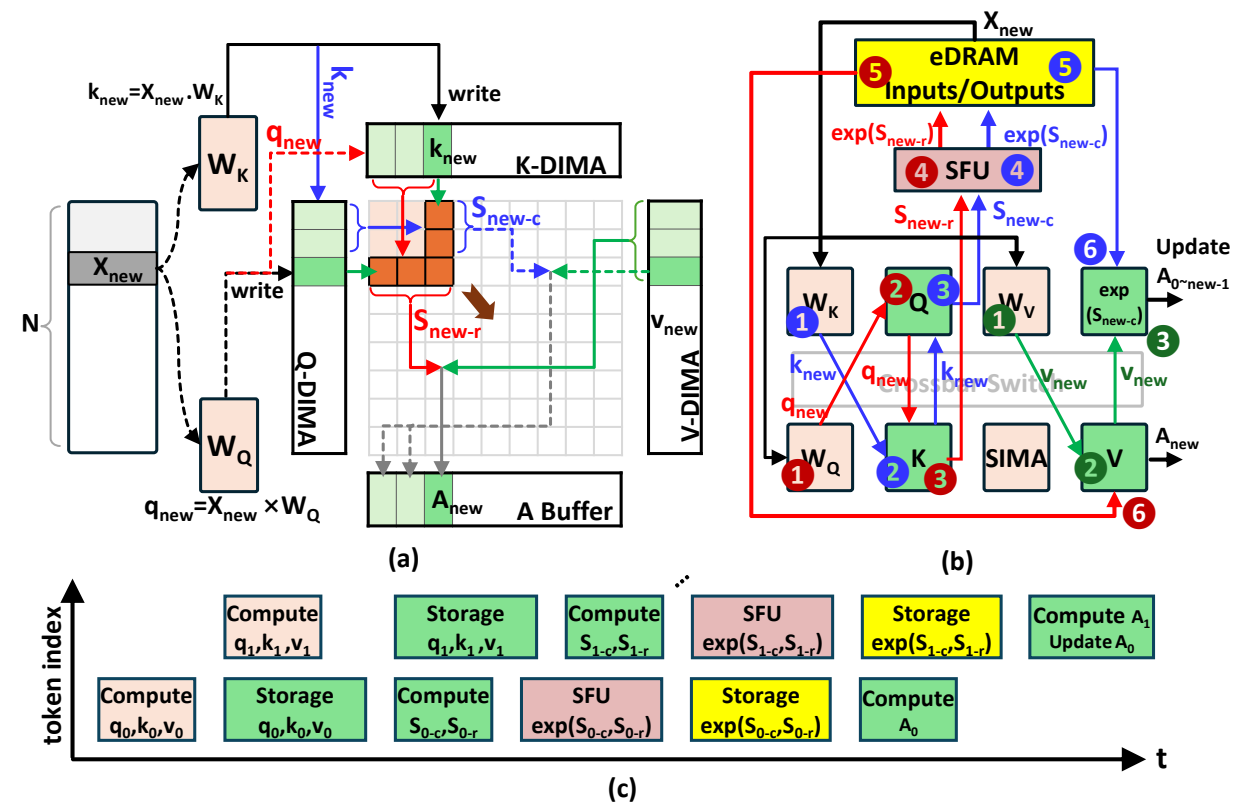

Fig. 5. IMC-friendly attention computation: (a) algorithm dataflow, (b) hardware dataflow, and (c) pipeline.

**Tile and Accelerator with Hybrid Memory:** A tile comprises eight IMAs, with half using SRAM for dynamic data processing and the other half using RRAM for static data processing. Dynamic and static IMAs (DIMA and SIMA) are interconnected through an internal crossbar switch, facilitating pipelined computing. Each tile includes a 128KB eDRAM as 8-bit inputs and outputs cache, a specialized function unit (SFU), and a quantization circuit includes 32KB memory. Like TIMELY, all tiles are interconnected via Hyper-Transport link. Upon completing a set of outputs, a tile can transfer data to other IMAs within the same tile via the internal I/O network or to IMAs of other tiles via on-chip network. Depending on the weight size, the accelerator can allocate one or multiple tiles to match the size of DNN layers.

### *D. IMC-Friendly Attention Computing Pipeline*

As transformer models gain prominence in the AI community, optimizing their computational flow becomes increasingly important. Fig. 5 illustrates an IMC-friendly attention computing flow in algorithm level. In this flow, static weights $W_Q$, $W_K$, and $W_V$ are pre-mapped into one or more SIMAs. The embedded word vector from eDRAM serves as the input vector for SIMA to generate *Q*, *K*, and *V* values for the current token, which are then loaded into DIMAs via an intra-tile crossbar switch. In *K-DIMA*, *K* is stored as weights in memory array, while *Q* is loaded into the input buffer and multiplied with all previously stored *K* values to get the score matrix. Conversely, in *Q-DIMA*, *Q* acts as the weights, and *K* is loaded into the input buffer to compute the scores with historical *Q* values. For each new input token, *K-DIMA* and *Q-DIMA* yields a $1\times n$ and $n\times 1$ score result ($S_{new\text{-}r}$ and $S_{new\text{-}c}$), respectively, where *n* denotes the current token position. Inspired by the flash attention approach, the newly generated score values undergo an exponential function transformation ($\exp(S_{new\text{-}r})$ and $\exp(S_{new\text{-}c})$) through the SFU and, along with the updated $l_{max}$ and $m_{ij}$, are stored in eDRAM. In subsequent cycles, these updated scores are multiplied by $V$ to refine the attention output. Once the last token has been processed, the complete attention output can be obtained immediately. Fig. 5(b) shows the hardware dataflow between components within a tile during attention computation, and finally, the computations between tokens can be organized in a pipelined manner to enhance the inference speed of the transformer (Fig. 5 (c)).

# IV. Evaluation

## A. Experiment Setup

TABLE II. Summary of YOCO Parameters

| | Compo. | Num. &Size | Energy (pJ) | Latency (ns) | Area/comp. ($\mu m^2$) |
|---|---|---|---|---|---|
| **MCC** | Capacitor | 2fF | 1.62fJ/act. | - | 0.8 |
| | SRAM/1T1R | 8/32 | - | - | 0.096 |
| **Array** | MCC array | 128×256 | [a]26.5 | 13 | 26214 |
| | Row driver | 128 | 9.36fJ | <30ps | 0.18 |
| | Time Acc. | 32 | 58.5fJ | 113ps | 5.3 |
| **IMA** | Array | 8×8 | [a]29.6 | <14.1 | 26406 |
| | [b]TDC (8bits) | 32×8 | 7.7 | 0.9 | 6865 |
| | I/O Buffer | 4KB | 2.9/256b | 0.112/256b | 4656 |
| **Tile** | IMA | 8 | 4325 | <15/VMM | 3.45$mm^2$ |
| | SFU | 128 | 0.6 | 0.1 | 1398 |
| | eDRAM | 160KB | 0.1pJ/bit | 128GB/s | *0.2*$mm^2$ |
| **Chip** | Tile | 4 | - | - | 27.8 $mm^2$ |
| **Total** | - | - | - | - | 111.2 $mm^2$ |
| **Hyper Link** | links/freq Line bw | 1/1.6GHz 6.4GB/s | - | - | 5.7 $mm^2$ |

[a]50% MCC activate; [b] Silicon-Verified by [8]

**YOCO Configuration.** All YOCO parameters, along with their energy, latency, and area, are summarized in Table II. Following the prior work [9], each MCC employs a 2 fF MOM capacitor for 8-bit computational resolution. The MOM capacitor can be stacked with memory cells, thus avoiding additional area in the layout. Multiple RAM units are organized as a memory cluster in MCC. To match the area of the MOM capacitor, each memory cluster in DIMA consists of 8 SRAMs, while in SIMA, each cluster contains 32 1T1R cells. Each array comprises 128×256 MCCs for 8-bit input and weight resolution, with each row driven by eight row drivers. Every eight columns of AiMC array form a compute-bar (CB), and the output of CB is connected to a column time accumulator. Each IMA consists of an 8×8 array, where analog computing results are read out in parallel by 32×8 TDCs. Both input and output data buffers for each IMA are set at 2 KB to enhance data reusability. Similar to ISAAC, YOCO integrates a special function unit (SFU) within each tile for complex function expansion and a 128 KB eDRAM for data caching.

**Methodology.** The YOCO model is constructed from multiple simulators or open-source data. Analog circuits, such as in-charge computing array and time-domain accumulator, are simulated in Cadence Virtuoso. TDC parameters are derived from previous report [10]. On-chip interconnects, SFU, buffer, and eDRAM are modeled using CACTI [11]. Parameters for ReRAM, featuring 1KΩ/20KΩ on/off resistance with 1-bit precision, are sourced from TIMELY [7]. Hyper-Transport links are based on ISAAC specifications [4]. All simulation data are then consolidated in the timeloop/accelergy framework, as referenced in REALLA [6], to perform overall architectural simulation. Given the 15ns computation latency of the IMA, the clock frequency for YOCO system simulation is set to 50MHz. Models of three baseline accelerators, ISAAC, TIMELY, and REALLA, are also developed in timeloop/accelergy [12]. For fairness, all accelerators are modeled at the 28nm process node, with shared components across designs.

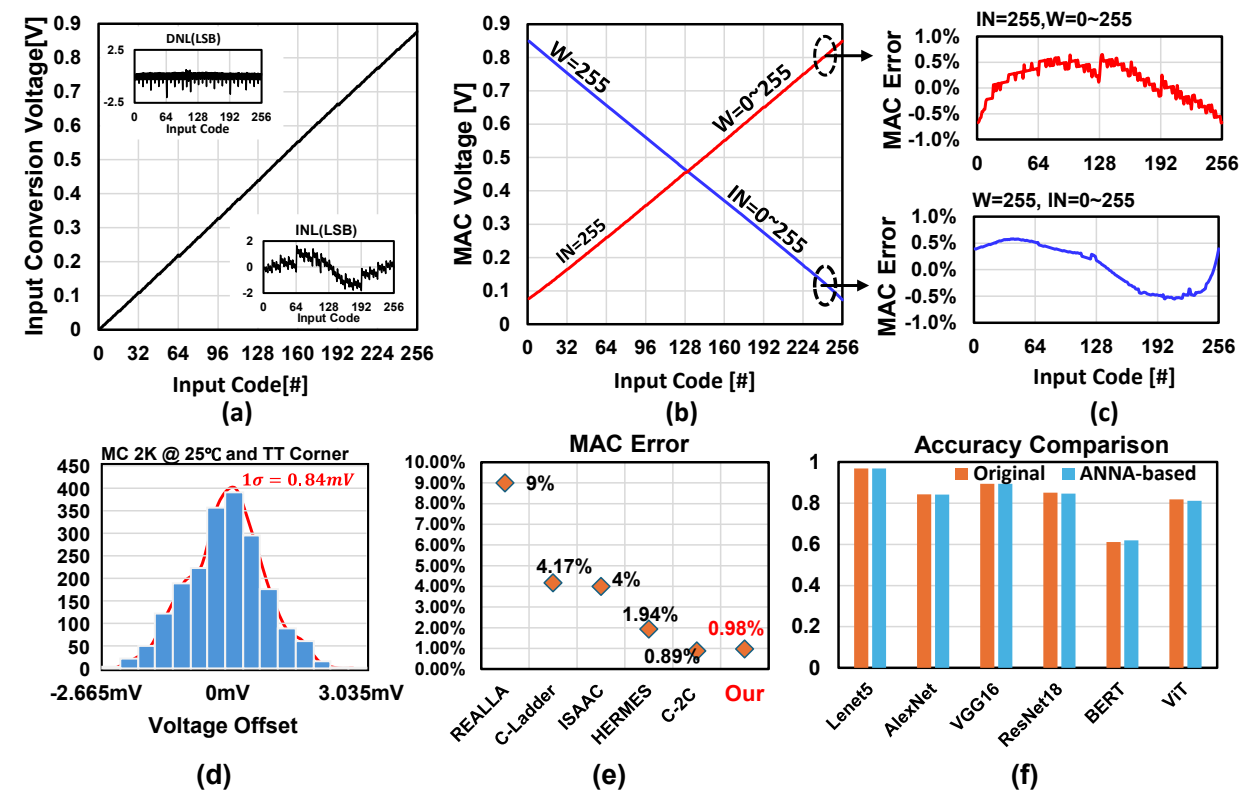

Fig. 6. An illustration of the (a) Transfer curve (TC) between input code and output voltage and correspond INL and DNL. (b) two TCs of 8-bit MAC with 128 input channels; and (c) correspond MAC error. (d) voltage offset under 2K Monte-Carlo simulation;(e) MAC error comparison; (f) DNNs inference accuracy comparison under 6 benchmarks.

**Benchmarks.** The overall YOCO architecture is evaluated over 10 DNN models, includes 5 CNN models (AlexNet, VGG16, ResNet18, MobileNetV3, and DenseNet201) and 5 transformed-based AI models (MobileBERT, QDQBERT, Vision Transformer, and LLaMA3-7B).

## B. Experiment Results

**Accuracy.** Fig. 6(a) shows the transfer curve (TC) between input binary code and conversion voltage, along with integral (INL) and differential nonlinearity (DNL). Conversion errors are within two least significant bits (LSB) and typically under one LSB. Fig. 6(b) provides two TCs during the 8-bit MAC operation with 128 channels: the blue curve represents the MAC voltage with weights scanned from 0 to 255 while input is fixed at 255, and the red curve shows input scanning from 0 to 255 with weight fixed at 255. The corresponding MAC error (less than 0.68%) is shown in Fig. 6(c). PVT variation analysis using 2,000 Monte Carlo simulations at TT corner and room temperature (Fig. 6(d)) indicates a 3σ variation of 2.25 mV—less than one LSB (3.52 mV). The time accumulator’s error is under 0.11%, keeping the analog MAC error below 0.79%. YOCO adopts 8-bit TDCs as the readout circuits for 8bit MAC resolution. The final end-to-end IMA computing error is less than 0.98% (see Fig. 6(e)). We also evaluated the impact of analog computing error for DNNs inference accuracy. Fig. 6(f) shows the accuracy comparison between fully-precision and YOCO-based inferences under 6 benchmarks. We observe less than 0.5% accuracy loss on 4 CNNs and less than 0.61% loss on 2 transformer-based models.

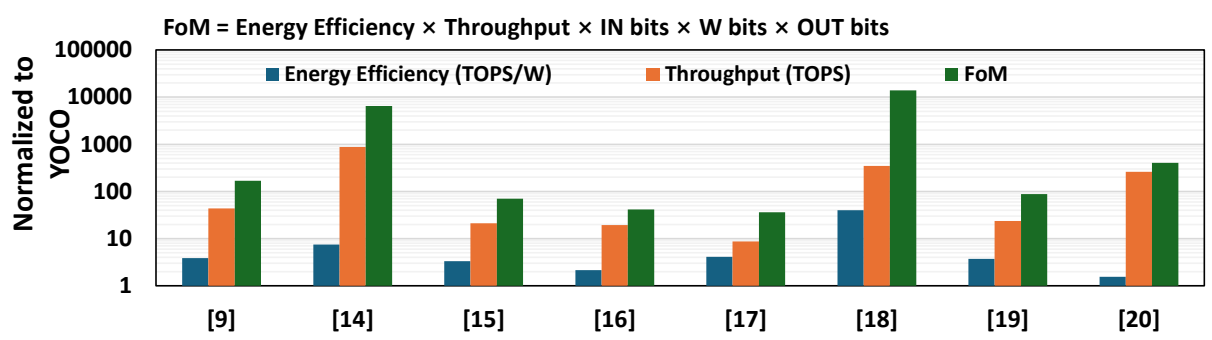

Fig. 7. The normalized VMM (a) energy efficiency and (b) throughput of YOCO's IMA over other SOTA design.

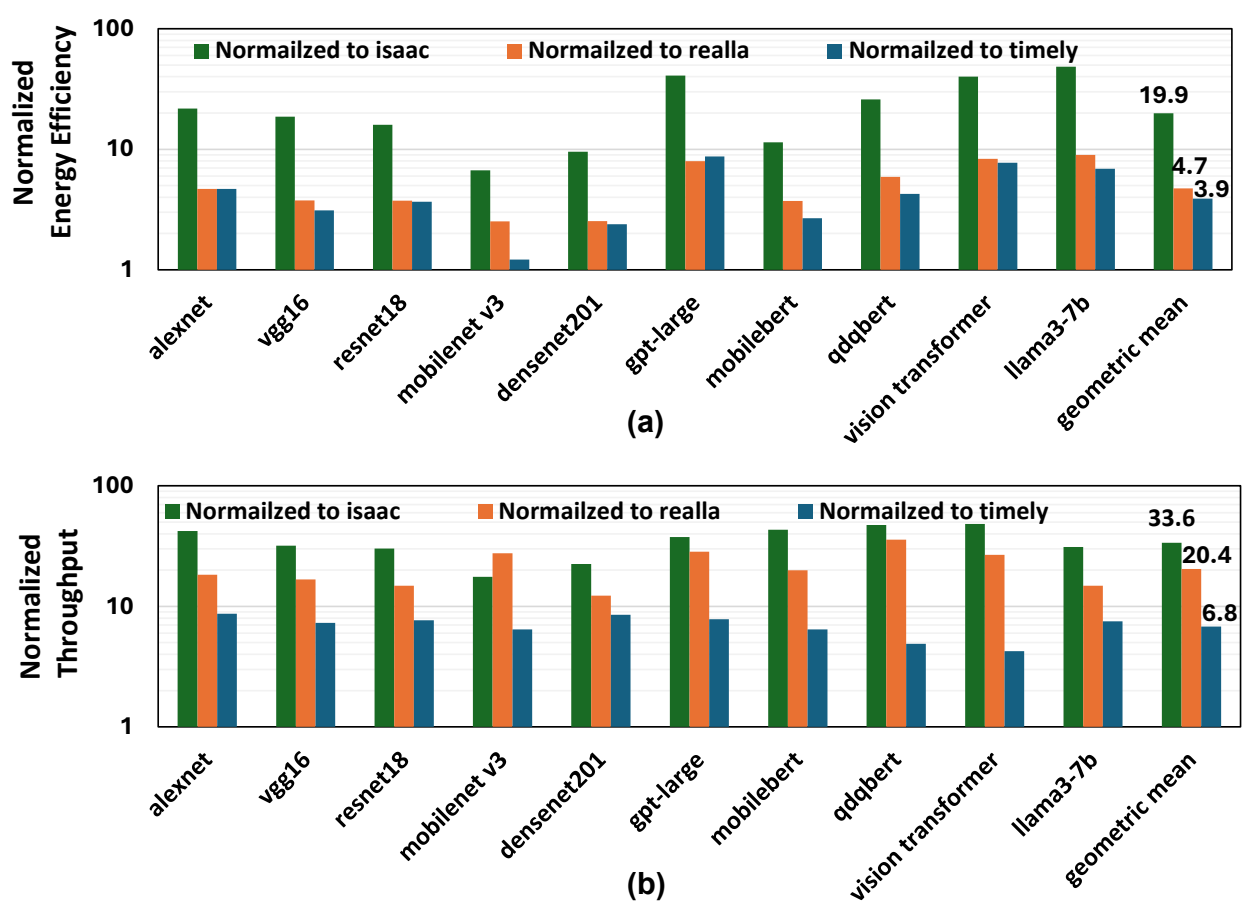

Fig. 8. (a) Efficiency and (b) throughput normalized to the ISAAC, REALLA, and TIMELY.

**Energy Efficiency and Throughput.** According to the summary in Table II, an IMA consumes approximately 4.235nJ of energy and incurs 15ns of latency to execute an 8-bit VMM with 1024×256 size. Based on discussion in [13], the average activation probability of MCCs is set to 50%. With these configurations and data, the IMA of YOCO achieves an energy efficiency of 123.8 TOPS/W and a throughput of 34.9 TOPS. Compared to 8 prior IMC circuits [9, 14-20], our design improves energy efficiency by 1.5× to 40×, throughput by 12× to 1164×, and the figure of merit (FoM) by 36× to 14,000× (Fig. 7). Furthermore, we evaluated the overall YOCO architecture on 10 CNN/transformer models to validate its superior performance. Fig. 8 shows the normalized energy efficiency and throughput over three state-of-the-art (SOTA) IMC architectures on 10 DNN models. On average, YOCO improves energy efficiency by 19.9×, 4.7×, and 3.9× and enhances throughput by 33.6×, 20.4×, and 6.8× compared to ISAAC, REALLA, and TIMELY, respectively (see the Geometric Mean in the rightmost part of Fig. 8).

**Save ADCs/DACs Cost:** Thanks to the innovative analog computing circuit design, YOCO significantly reduces the costs associated with expensive DACs and ADCs, while achieving ultra-high energy-efficiency and throughput. For DACs, YOCO uses grouped row capacitors to replace 8bits DAC, enabling low-cost input data conversion. Fig. 9(a) compares the DAC overhead between traditional and YOCO's methods, showing that YOCO reduces DAC area by 352×, energy consumption by 9×, and latency by 1.6×. For ADCs, YOCO adopts novel charge-domain computation and time-domain accumulation approaches to drive an all-analog 8-bit MAC, minimizing the number of ADC conversions while maintaining accuracy. As shown in Fig. 9(b), compared to bit-wise input , YOCO reduces the area, energy, and delay of ADC by 98.4%. When compared to weighted in digital-domain schemes, YOCO reduce ADC area and energy costs are reduced by 87.5% without delay cost.

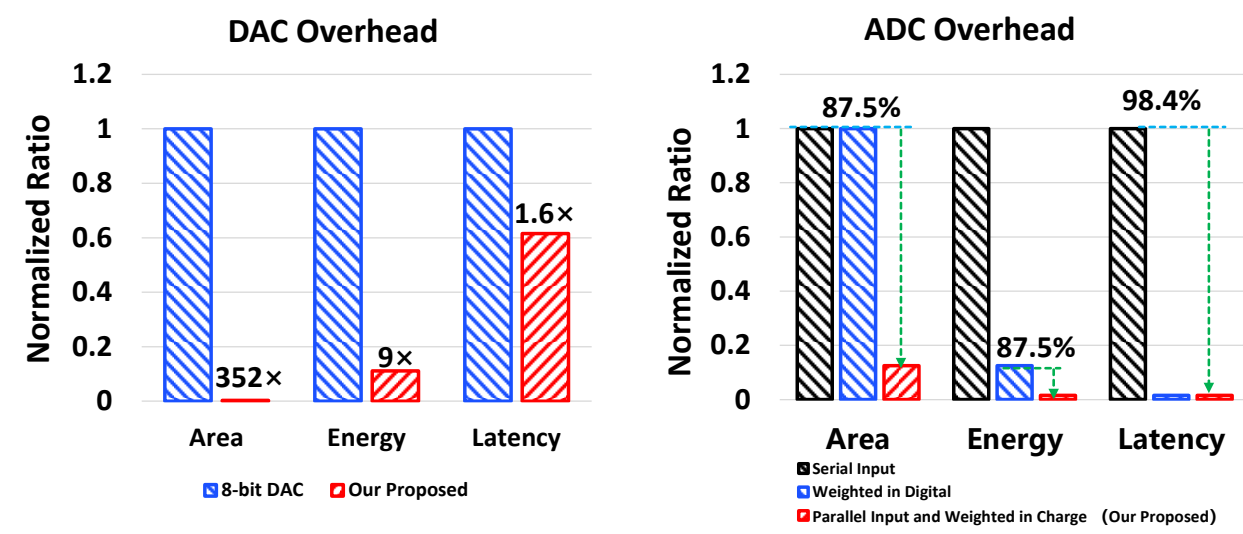

Fig. 9. (a) DACs and (b) ADCs overhead comparison with traditional data conversion strategies.

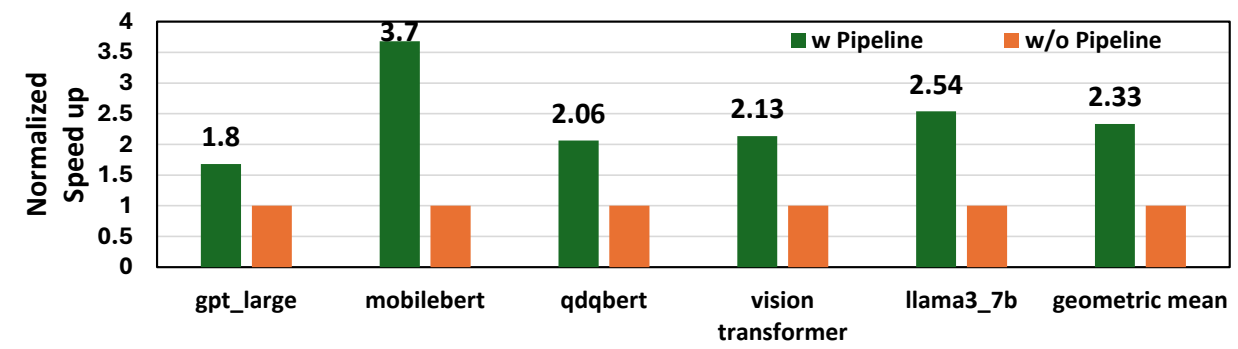

Fig. 10. The normalized speed up with or without pipeline design for 5 transformer models' inference.

**Pipeline Performance.** We also evaluated the effectiveness of the proposed IMC-based pipelined computation for accelerating attention layer. In 5 classic transformer models, compared to traditional layer-wise computation dataflow, the pipelined approach achieves speedups of up to 1.8× to 3.7× (geomean 2.3×) for attention inference (Fig. 10). The performance gains primarily stem from effectively hiding the computing latency of attention layer.

## V. Conclusion

YOCO is the first work to demonstrate energy efficiency exceeding 100 TOPS/W in 8-bit fixed-point arithmetic. This breakthrough significantly alleviates the energy consumption bottleneck faced by large-scale model computations. YOCO marks the first successful demonstration of the superiority of analog-based computing. Benefits from hybrid memory architecture and optimized attention computation pipeline, YOCO is also uniquely positioned to tackle the rapidly evolving demands of large language models (LLMs). These innovations ensure that YOCO delivers both high performance and energy efficiency, making it a cutting-edge solution for future AI.

## Acknowledgments

We sincerely thank Dr. Jia Chen for her valuable support and guidance throughout the manuscript submission process.